\documentclass[notitlepage, twoside, a4paper, twocolumn]{article}
\usepackage[margin = 2cm, top = 2cm, bottom = 2cm]{geometry}
\usepackage{titling}
\usepackage{lipsum}
\usepackage[utf8]{inputenc}
\usepackage{authblk}
\usepackage[english]{babel}
\usepackage{multicol}
\usepackage{amsmath}
\usepackage{appendix}
\usepackage{graphicx}
\usepackage{cite}
\usepackage{float}
\usepackage{comment}
\usepackage{xcolor}
\usepackage{hyperref}
\usepackage[squaren]{SIunits}

\title{\vspace{-1.9cm}\textbf{\Large{Zero-power calibration of photonic circuits at cryogenic temperatures}}}
\let\oldbibliography\thebibliography
\renewcommand{\thebibliography}[1]{%
  \oldbibliography{#1}%
  \setlength{\itemsep}{0pt}%
} 
\addto\captionsenglish{}
\author{\normalsize{Ben M. Burridge$^{1, 2}$, Gerardo E. Villarreal-Garcia$^{1}$, Antonio A. Gentile$^{1, 3}$, Pisu Jiang$^{1}$, Jorge Barreto$^{1*}$}}
\affil{\small{$^1$Quantum Engineering Technology Laboratories, University of Bristol, Bristol, United Kingdom.\\ $^2$Quantum Engineering Centre for Doctoral Training, Centre for Nanoscience \& Quantum Information, University of Bristol, Bristol, United Kingdom.\\
$^3$ Qu \& Co BV, Amsterdam, the Netherlands.\\
$^*$ Corresponding Author: g.barreto@bristol.ac.uk}}
\date{}

\begin{document}
\onecolumn
\maketitle

\begin{abstract}
\textbf{The continual success of superconducting photon-detection technologies in quantum photonics asserts cryogenic-compatible systems as a cornerstone of full quantum photonic integration. Here, we present a way to reversibly fine-tune the optical properties of individual waveguide structures through local changes to their geometry using solidified xenon. Essentially, we remove the need for additional on-chip calibration elements, effectively zeroing the power consumption tied to reconfigurable elements, with virtually no detriment to photonic device performance. We enable passive circuit tuning in pressure-controlled environments, locally manipulating the cladding thickness over portions of optical waveguides. We realize this in a cryogenic environment, through controlled deposition of xenon gas and precise tuning of its thickness using sublimation, triggered by on-chip resistive heaters. $\pi$ phase shifts occur over a calculated length of just $L_{\pi}$ = 12.3$\pm$0.3~$\mu m$. This work paves the way towards the integration of compact, reconfigurable photonic circuits alongside superconducting detectors, devices, or otherwise.}
\end{abstract}

\section{Introduction}

Photonic integrated circuits (PICs) offer a promising glimpse at the prospect of universal quantum information processing using quantum states of light\cite{o2009photonic,qiang2018large, Obrien2007opticalqc,silverstone2015qubit,wang2019integrated,silverstone2016}. 
As we progress towards a fully integrated, reconfigurable quantum photonic platform, single-photon detectors must be implemented alongside existing devices to remove bottlenecks in scaling to larger circuits, minimize feed-forward associated delays, and mitigate optical coupling losses.
Current best-in-class detectors, superconducting nanowire single-photon detectors (SNSPDs) \cite{doi:10.1063/1.3657518,Reddy:19} combine outstanding performance metrics in dark count rates, efficiencies, and jitter time. Furthermore, the realization of SNSPDs on integrated photonic circuits facilitates a further boost to their performance \cite{Pernice_2012,tyler2016modelling,reithmaier2013chip,sahin2013waveguide,schuck2016quantum,schuck2013waveguide,najafi2015chip}, enabling a scalable approach at the cost of low-temperature (typically below 4 kelvin) operation.

Modern reconfigurable PICs are reliant on methods of static compensation to maximize overall performance and ensure the proper functionality of PICs\cite{Bogaerts2020}. Phase modulators, such as thermo-optic \cite{Watts:13,silverstone2014chip}, electro-optic \cite{8613782} and plasma-dispersion based devices \cite{Reed2010} are often used at ambient temperature. Historically, the basic levels of integration achieved have allowed the constraints of these particular devices to be somewhat compensated for or overlooked. Reconfigurable PICs were thus successfully used to explore and demonstrate small scale quantum experiments with high fidelity \cite{wang2017experimental,Faruque:18,llewellyn2020chip}. Although thermo-optic phase shifters (TOPS) have been successfully tested in cryogenic environments \cite{7463458}, power requirement concerns render them infeasible for large-scale photonic circuits within cryogenic systems. On the other hand, the use of exotic non-linear materials such as barium titanate allows for efficient high-speed electro-optic modulation at cryogenic temperatures as reported recently \cite{eltes2020integrated}.

Here we demonstrate a method of tuning the optical properties of an individual waveguide element that breaks with convention. We facilitate the deposition of a xenon (Xe) cladding on the surface of an exposed waveguide core by controlling environmental pressure and temperature. We demonstrate this method's functionality using a Silicon-On-Insulator (SOI) photonic circuit - containing a Mach-Zehnder Interferometer (MZI) with one of its arms exposed to our controlled environment. We take full advantage of the cryogenic conditions required for the operation of SNSPDs to dynamically and reversibly adjust the geometry of exposed sections of waveguide. We also re-purpose TOPS as on-chip temperature controllers. They function as localized and precise heat sources capable of sublimating the deposited cladding material, thus controlling the effective refractive index.

Our work expands on the ideas presented previously by \cite{Mosor2005ScanningXenon} and \cite{li2015coherent}, aimed at the control of resonances in optical nano-cavities. We use localized on-chip heat delivery to address specific circuit elements and control the structural properties of individual components. Once the devices are operating as required, the waveguide components retain these characteristics as long as the environmental conditions are kept stable. As a result, cladding layer manipulation (CLM) requires no static power, releasing the entire on-chip power budget to more crucial components.


\section{Theoretical Background}

The power budget of a PIC is an often-overlooked parameter of current-generation photonic devices. However, modern cryostats have limited cooling powers (typically on the order of a fraction of a watt at temperatures close to 1 kelvin), which can be quickly saturated by power-hungry devices. PICs must therefore aim to function optimally in these conditions. 
We can separate the power consumption of integrated devices into two distinct categories based on their role within a PIC, namely controllable and configurable components. Controllable components serve to rapidly re-arrange the optical paths of a circuit into specific configurations that are, for example, capable of processing information. An example of this is re-routing a photon, based on a feed-forward signal \cite{kaneda2019high}.
Conversely, configurable components program and stabilize the optical path for consistent operation; with the added benefit of reconfigurability as required. Components of this kind can therefore compensate for inaccuracies in the PICs stemming from the fabrication process\cite{articleb}. For example, waveguide structures such as directional couplers (DCs), and to a lesser extent multi-mode interferometers (MMIs) \cite{296191} are sensitive to fabrication tolerances \cite{Rajarajan:99}. These imperfections may ultimately translate to infidelities when preparing quantum states using the PIC. MZIs combine either DCs or MMIs, alongside a method of phase control, and are regarded as a tool for post-manufacture reconfiguration \cite{silverstone2015qubit,silverstone2016,wang2019integrated,Bogaerts2020}. However, this comes at the cost of an increase in circuit complexity and static power consumption. 

Waveguides on the SOI platform confine and guide light using the contrast between silicon's (Si) high refractive index (3.476 at 1550~$nm$) that forms the waveguide core; and the lower index material of silica (SiO$_2$, 1.44 at 1550~$nm$) that creates the waveguide cladding. In standard waveguide geometries (500~$nm$ x 220~$nm$), the optical mode is mostly present in the waveguide core, with a small fraction of evanescent mode propagating in the cladding. The cladding layer acts as a protective barrier to the waveguide core, diminishing the interaction between the externally propagating evanescent field and the environment. Consequently, the composition of these layers and their refractive index strongly affect the supported optical modes.


The effective refractive index determines how much phase an optical mode accumulates as it propagates, influenced by both the waveguide core and the cladding layer. The resultant optical path stems from the interference of all the possible optical paths. Therefore, we can alter the refractive indices of either material to control the final path of the mode. Prevalent techniques target the optical properties of the waveguide core as it carries most of the propagating mode \cite{Watts:13,Reed2010}.

\begin{figure}[t]
    \centering
    \includegraphics[width = \linewidth]{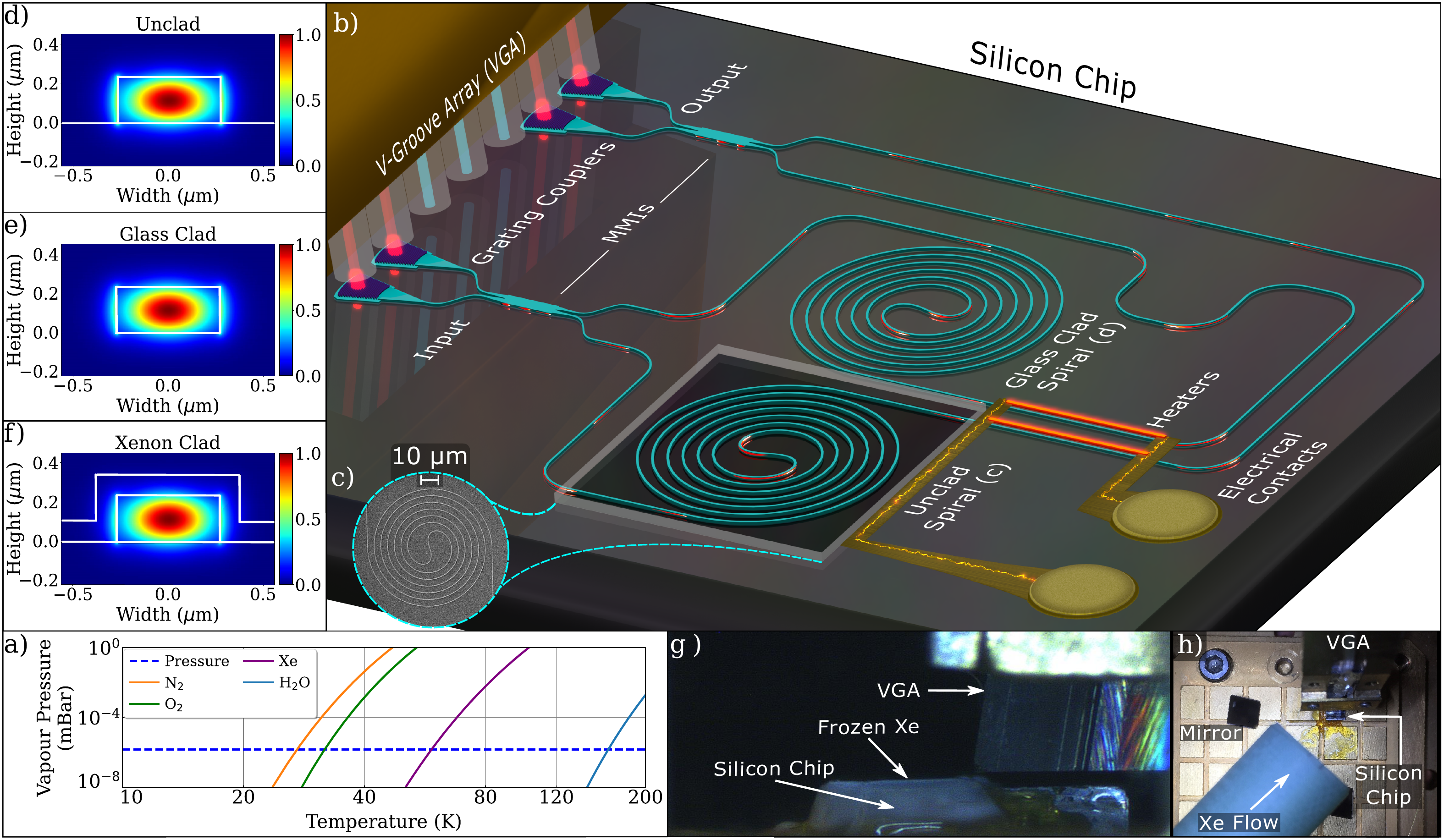}
    \caption{Simulation results and experimental design. \newline a) Vapor pressures of relevant gases \cite{honig1960vapor}. The dashed line highlights the chamber's base pressure. b) Artist's impression of the photonic circuit. Light is injected and collected using an array of optical fibers coupled to input/output grating couplers. MMIs are used to split the light 50:50, forming the interferometer. The electrical contacts are used to apply a voltage to the heater elements, placed above the cladding close to the long spiral delay arms. c) Scanning Electron Microscopy image of the exposed 1056$\pm$2$~\mu m$ section of waveguide, with 10~$\mu m$ scale; d - f) Simulated mode profiles for unclad, glass and ideal Xe clad waveguide cross-sections. g) Image capture of the experiment after a full-film deposition with Xe build up on the chip. h) An overhead view of the experimental setup, mirrors were used to align the setup while under vacuum.}
    \label{fig:Rendering}
\end{figure}

The cladding material plays a comparatively minor role via the evanescent field. However, it is possible to precisely modify the effective refractive index the optical modes see by shaping the cladding layer. 
Under specific environmental conditions, substances can accumulate and form a film on the device surface via condensation or deposition. We can harness these phenomena by controlling the temperature, pressure, or chemical composition of the environment.
For example, in a pressure-controlled chamber, a gaseous substance will condense (deposit) on a surface provided that the substance's vapor pressure at the surface temperature is below the chamber pressure. 
Temperatures and pressures where this can occur are determined using the gas' vapor pressure curve \cite{honig1960vapor} (as reproduced in Fig.~\ref{fig:Rendering}a). Conversely, a solid substance at a given temperature will sublimate (desorb) if its vapor pressure is higher than the pressure exerted by its environment. A chamber pressure in equilibrium with substance vapor pressure results in a balancing of condensation and sublimation rates, leading to an inherent lack of control over the deposited film.

Accurate on-chip sensing of waveguide, and specifically cladding changes induced by the presence of thin solid films is possible using an MZI \cite{fabricius1992gas}. Relative differences in accumulated phase; for instance due to different waveguide geometries, between the two arms of the MZI results in a significant interference pattern at the output, providing an easy map between phase and intensity. Long waveguide sections may be used as a sensor, by taking advantage of their large interaction region with the environment (Fig.~\ref{fig:Rendering}b-c).

The free spectral range (FSR) is the wavelength separation between successive peaks (or troughs) in the transmission of an unbalanced interferometer setup. The FSR of an MZI can therefore be used to extract the difference in effective group index ($n_{g}$) between two equal length arms of the interferometer, using equation 1. 

\begin{equation}
    FSR = \frac{\lambda^{2}}{\Delta n_{g}L}
    \label{FSR Eqn}
\end{equation}

\noindent Here $\lambda$ is the central wavelength used to measure the FSR of the MZI, $L$ is the length of the exposed spiral, and $\Delta n_{g}$ is the change in effective group index between the exposed and fully clad waveguide arms of the MZI.

We follow the experiment by Mosor \cite{Mosor2005ScanningXenon} and use Xe as our substance of choice. Xe is non-reactive, has a reasonably high vapor pressure at low temperatures, and a solid phase refractive index of 1.47\cite{GRACE2017204} close to that of an SiO$_2$ cladding. Temperatures lower than the deposition temperature of oxygen (O$_2$) and nitrogen (N$_2$) require higher vacuums or temperature cycling of the device to minimize contamination stemming from the unwanted deposition of any other gases in the environment.

\section{Materials and Methods}

\subsection*{Fabrication}

We designed an integrated Mach-Zehnder Interferometer (MZI) on a commercially-available SOI platform (Fig.~\ref{fig:Rendering}b) to characterize the modulation capability of Xe. We used standard strip waveguides of 500~$nm$ width x 220~$nm$ height for operation within the telecom C-band and to support the fundamental TE mode. The SOI platform uses a 2$um$ layer of SiO$_{2}$ buried oxide (BOX) and another 3$um$ of SiO$_{2}$ cladding. Thin metallic layers (2~$\mu m$ - aluminium and 120~$nm$ - titanium nitride respectively) are deposited and patterned on top of the cladding to form electrical wires, pads and resistive heaters to be used for localized thermo-optic phase shifting over specific waveguide sections.
The SiO$_{2}$ cladding was removed from one of the arms of the MZI using a Buffered Oxide Etch (BOE 7:1), exposing the spiral section entirely.

\subsection*{Simulation}

Simulations (Lumerical MODE) calculated the modes supported by the cross-section of a straight waveguide (Fig.~\ref{fig:Rendering}d-f). We approximately modelled how Xe would deposit on top of the waveguide and assumed structural symmetry because of its spiral nature (Fig.~\ref{fig:Rendering}c). The waveguide's geometry will ultimately affect how Xe distributes across it. Surfaces without line-of-sight to Xe flow (shadowed) would see reduced deposition; conversely, direct line-of-sight implies enhanced growth. Simulated Xe shadows are at an angle of 60$^\circ$ to the horizontal, approximated from the experimental setup shown in Fig.~\ref{fig:Rendering}(g-h). We also make an assumption based on the Knudsen number, (the ratio between the mean free path of Xe \cite{jousten2018handbuch} and the vessel diameter) that Xe flows in the molecular regime at the low pressures of our cryogenic vessel. To emulate an over-etch, we removed SiO$_{2}$ from under the waveguide until the group indices of experiment and simulation coincided. Within the experimental error, this occurred for undercuts of 200$nm$. The thermo-optic effect plays a large role when dropping from room temperature to cryogenic temperatures. Therefore, we extracted the group indexes of Si \cite{frey2006temperature} and SiO$_{2}$ \cite{leviton2006temperature} from refractive index data around cryogenic temperatures. Similarly, the group index of Xe was extracted using its extrapolated refractive index \cite{GRACE2017204}. 

\subsection*{Circuit design}

We used grating-couplers as optical inputs and outputs to interface with the chip. These are standard components optimized for the quasi-TE mode and an angle of incidence of 11$^{\circ}$ to avoid back reflections from second-order diffraction effects \cite{Taillaert_2006}. 
The input light is split using a multi-mode interference structure (MMI), acting as balanced beam-splitters designed for a 50:50 splitting ratio. We used MMIs for their relative resilience to fabrication imperfections \cite{296191} and to reduce the number of variables we had to control during the experiment. Waveguide sections of 1$mm$ in length were arranged in a spiral configuration and placed on each side of the MZI; both arms were then path-matched.
The MZI structure is completed with a second 50:50 MMI and coupled out of the chip into optical fibers through two additional grating couplers. 

Heater elements (each 180$\mu m$ in length) are located symmetrically along each arm of the MZI to provide a local heat source. These follow standard design rules used for TOPS. Both heater elements are connected electrically in parallel (centered symmetrically within 170~$\mu m$ of the exposed region) to minimize any phase-shift in the optical path induced by heat. The dimensions of the cladding openings (windows) limit heater proximity to the exposed waveguides. This makes it difficult to estimate accurately the amount of heat dissipated by the resistor that reaches the waveguide sections. We give the total power consumption ($P_{Total}$) for every example, but only half of that ($P_{Local}$) is dissipated in immediate proximity to the spirals; the other half is at a distance of $400\mu m$. 

\subsection*{Experimental setup}

We placed the device under test (DUT) inside a continuous-flow cryogenic probe station (Lakeshore CPX). We ensured thermal contact with the probe station sample-stage using a high thermal conductivity varnish (CMR-Direct GE-7031). The probe station was evacuated to 1$\times10^{-4}$ $mBar$ using an Agilent TPS-compact pumping system, and then cooled using liquid helium. Intrinsic system cryopumping brought the internal pressure down to 1$\times10^{-6}$ $mBar$. The sample stage temperature was stabilized by adjusting the helium flow rate and local heating using PID temperature controllers (Lakeshore Model 336 Temperature Controller). We operate between the equilibrium vapor pressure of Xe and O$_2$ at our chosen range of temperatures, reducing the likelihood of any external contamination while maintaining negligible sublimation rates of Xe. 

We controlled the Xe flow rate using a Bronkhorst F-201CV mass flow controller (MFC) connected to a Xe line pressurized to $>$1 $atm$ to prevent external contamination. The Xe used is 99.999\% pure, and the lines between the gas bottle and the main chamber were evacuated to 1$\times10^{-4}$ $mBar$ before the first pressurization. We directed Xe flow using a bi-axial translation stage connected to a wobble stick, with the Xe entering the main chamber through a nozzle of diameter 2$mm$ (Fig.~\ref{fig:Rendering}h).

\subsection*{Measurement procedures}

We used custom software to control the experimental hardware and synchronize data acquisition, in addition to real-time monitoring of the Xe's effect on the DUT.

The DUT was probed optically using a C-band tunable CW laser source (Yenista T100S-HP centred at 1550$nm$), paired with a Yenista CT440 to accurately set the wavelength of the laser and collect full optical spectra (1510$nm$ - 1590$nm$) from both outputs of the DUT. We set the input light polarization using a strain-based polarization controller, maximizing light coupled into the PIC.

Once the vacuum chamber reached its base temperature, we lowered the sample stage temperature to ranges determined from the data in Fig.~\ref{fig:Rendering}a. We coupled light into and out of the chip through the use of the on-chip grating couplers and used the balance of optical power between the different outputs of the on-chip MZI as a means to observe the deposition and sublimation processes.

The sample temperature was kept at 50-43~$K$ while lowering the external radiation shields to 4.8~$K$, 10~$K$, and 19~$K$. The sample stage temperature target was determined using Fig.~\ref{fig:Rendering}a. This data was refit using the Antoine equation \cite{doi:10.1021/cr60119a001}; optimized to be accurate over temperatures where vapor pressure lines intersect experimental vessel pressures. Here, we chose 50-43~$K$, settling on 45-43~$K$ for the longest measurements due to subtle sublimation observed at 50~$K$. Xe was injected into the chamber in discrete steps, with a flow range between 3 ml/min and 15 ml/min. Flow rates and injection times were changed dynamically throughout the experiment as the Xe films saturated.
Once Xe had deposited on the waveguide surface, we used on-chip TOPS to accurately control the local temperature and sublimate the Xe at a controlled rate. Xe sublimation was tested using two pre-programmed depositions, followed by controlled sublimation until no further change was observable. Multiple films were deposited for consistency, and we observed no noticeable changes in film deposition behavior.

\section{Results}

\subsection*{Simulations}

We simulate the optical modes supported by the device without any cladding (Fig.~\ref{fig:Rendering}d), with a conventional glass cladding (Fig.~\ref{fig:Rendering}e), and for different Xe cladding layer thicknesses (Fig.~\ref{fig:Rendering}f). We then extract the simulated values of $n_{g}$ to estimate the change in Xe layer thickness (Fig.~\ref{fig:Ng_with_Vol}) of our MZI test device (Fig.~\ref{fig:Rendering}b).

Our model accounts for a potential under-etch during the fabrication process by including a 200$nm$ undercut in the buried oxide (BOX) layer. Additionally, we consider variations in waveguide height and width according to deviations shown in \cite{lu2017performance}. Initial simulations estimate the group index to be 4.35, rather than the experimentally observed 4.43 (Fig.~\ref{fig:Ng_with_Vol}). This value can be simulated more accurately if a certain undercut is assumed. Finally, we also account for the apparent directionality of Xe flow (Fig.~\ref{fig:Rendering}(g-h)). Our simulations suggest that a solid Xe film of thickness 200~$nm$ will produce a considerable phase shift in the MZI, by reducing $n_{g}$ from 4.43 to 4.27. 

Analysis of the modal overlaps between the clad and unclad sections of the undercut waveguide predicts total device insertion losses of $\leq$0.0226~$dB$. Specifically, the mode mismatch between SiO$_2$ and vacuum (or air) interfaces result in scattering losses of 0.0113 $dB$. On the other hand, the interface between a waveguide clad in solid Xe (1~$\mu m$ thickness), and SiO$_2$ lowers the modal mismatch and associated loss to 0.0064~$dB$. Our simulations suggest that ideal (non-undercut, as in Fig.~\ref{fig:Rendering}f) Xe geometries would have mismatch losses of as low as 0.0002~$dB$ for saturated layers (additional Xe induces no further change). Device insertion losses such as these can be effectively neglected when we make comparisons to record low waveguide propagation losses of 2.7~$dB/m$ \cite{biberman2012ultralow}.

\subsection*{Deposition}

\begin{figure}[h]
    \centering
    \includegraphics[scale = 0.3]{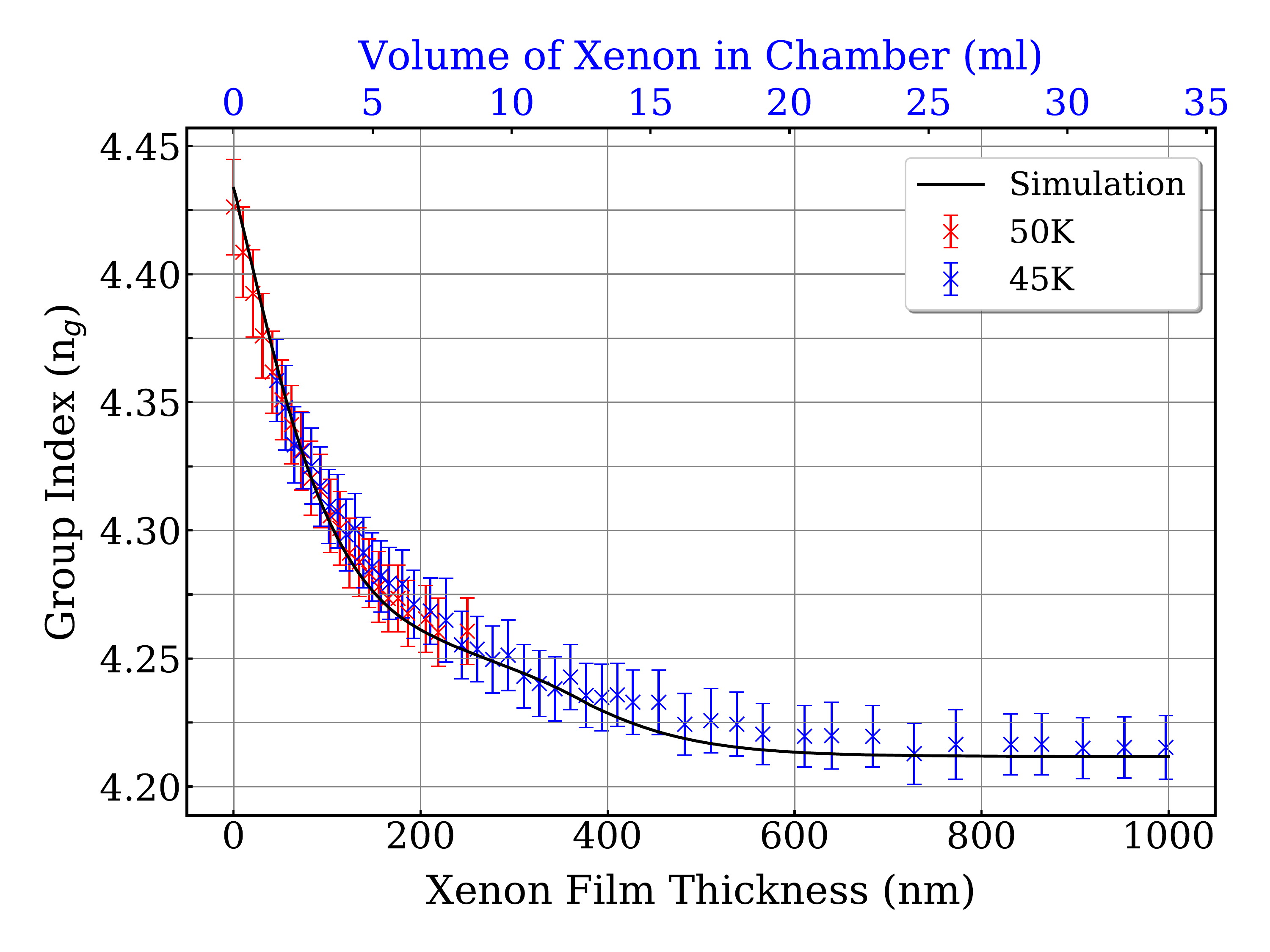}
    \caption{Dependence of the group index on Xe film thickness. \newline The experimentally extracted values of group index ($n_{g}$) with the volume of Xe introduced into the chamber. The black line represents the simulated results of our model, calculating the corresponding $n_{g}$ with Xe layer thickness. This is our map between experimental $n_{g}$ and the simulated thickness of a Xe film.}
    \label{fig:Ng_with_Vol}
\end{figure}

We monitor the power balance of the on-chip MZI to observe real-time shifts in the relative phase of the two arms during Xe deposition. Further, we extract the FSR of the structure by spectrally scanning
our laser source to measure the wavelength of successive peaks in transmission. In the general case, we map the FSR values to our simulated data via $n_{g}$ (Fig.~\ref{fig:Ng_with_Vol}) using equation 1, ultimately providing us with an experimental estimate of Xe layer thickness.

We designed both arms of the MZI to be the same length, with the unclad waveguide length estimated to be $1056\pm 2\mu m$; the MZI is therefore unbalanced with an FSR of $8.3\pm0.1~nm$.

The change in $n_{g}$ with Xe thickness is less significant as the film grows, and it becomes negligible (saturated) beyond 700~$nm$-thick films. Xe's effect gradually diminishes due to the increase in distance from the waveguide core, effectively reducing the interaction with the evanescent component of the mode propagated along the waveguide.

We used higher deposition rates to reach deposited film thickness values above 400~$nm$. An FSR of $36.2\pm1.2~nm$ was measured for saturated Xe films, equivalent to a SiO$_2$-clad path mismatch of 15.8~$\mu m$, or 15~$nm$ per micron of the exposed waveguide. 

\subsection*{Sublimation}
Once Xe has been deposited, we use on-chip heaters (see Fig.~\ref{fig:Rendering}b) to increase the temperature in the vicinity of the exposed waveguide section. We control the temperature via the power dissipated resistively, and the total energy delivered depends on the duration of the electrical pulse. 
We monitor the power output of the MZI and use interference data obtained during deposition to characterize the sublimation of the Xe films (Fig.~\ref{fig:Ng_with_Vol}). Plotting the oscillation in optical power out of the chip allows us to observe this accumulation of phase, which we wrap to the phase difference ($\Delta\phi$) between the two arms of the MZI.

\begin{figure}[h]
    \centering
    \includegraphics[width=\linewidth]{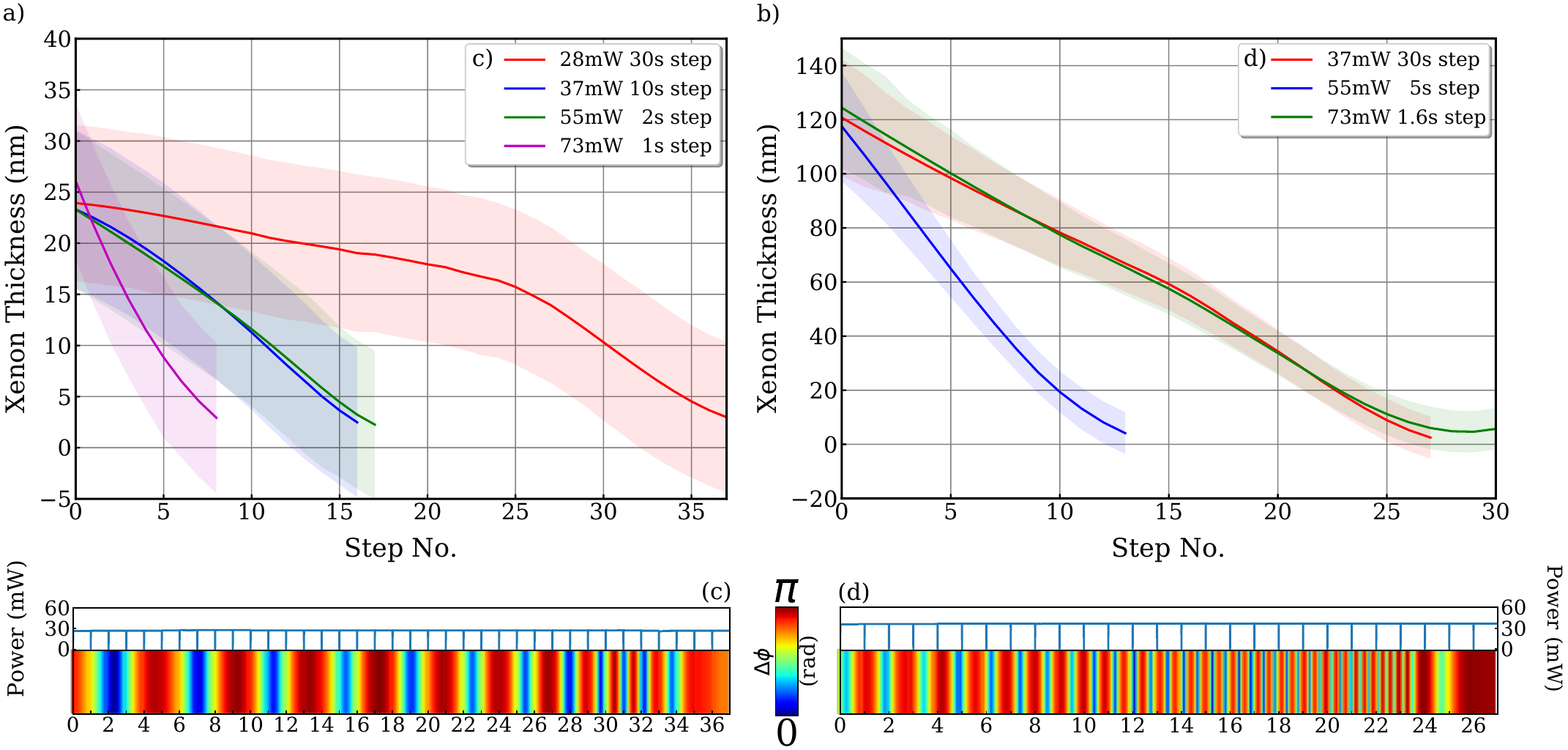}
    \caption{Controlled local Xe sublimation using resistive heaters. \newline Extrapolated values for Xe film thickness following sequential heater cycles. The translucent regions indicate the uncertainty.
    The heaters were activated at set power levels, and for set time periods (steps) to illustrate the sublimation process obtained starting from either, a) Thin (23.9$\pm$7.7 $nm$) or b) Thick (120.8$\pm$21.4 $nm$) Xe films.
    Plots (c, d) below each graph depict the electrical power delivered and the oscillation in relative phase ($\Delta\phi$) between the two arms of the MZI as the Xe sublimates with time. $\Delta\phi$ is approximately mapped from the optical power at one of the outputs of the chip.}
    \label{fig:Sublimation}
\end{figure}

Figure \ref{fig:Sublimation}(a-b) shows the thickness extrapolated from FSR measurements for different sequences of heater steps of specific power and time.
For context, we observed no sublimation of thin Xe films (23.9$\pm$7.7~$nm$ thickness) when addressing them with power cycles dissipating a total power ($P_{Total}$) of $<18.8$ $mW$. As a result, this suggests induced waveguide temperatures of $<$60~$K$. 
We also deposited thick films (120.8$\pm$21.4~$nm$ thickness, Fig.~\ref{fig:Sublimation}b) to scrutinize any potential changes in behavior compared to thinner films. Rates of sublimation can be approximated with a linear response for heater powers above $P_{\text{total}}$ = 37 $mW$, for both thick and thin films. As can be seen in Fig.~\ref{fig:Sublimation}, a 2-fold increase in power typically results in an order of magnitude reduction in the time required to completely sublimate a thin film.

Figures ~\ref{fig:Sublimation}c \& d show both the measured phase and the power delivered with each consecutive step. The phase oscillates with gradually decreasing period as the Xe films become thinner.
The cumulative total phase change ($\Delta\phi$) for thin films is 24.84$\pm$0.04 $\pi$; thick films demonstrated a total $\Delta\phi$ of 71.5$\pm$0.5 $\pi$, and saturated depositions ($>$600$nm$) exhibited a total $\Delta\phi$ of 86.0$\pm$2.0 $\pi$ with a predicted $\Delta\phi$ of 85.0 $\pi$. For context, if we had used TOPS; like those re-purposed to sublimate Xe (Fig.~\ref{fig:Rendering}b), to achieve 86 $\pi$, the device would have a length of over 8 $mm$. This is calculated using the approximate waveguide temperature \cite{jacques2019optimization} at the highest powers used here (73 $mW$, Fig.~\ref{fig:Sublimation}a-b).

\begin{figure}[h]
    \centering
    \includegraphics[scale = 0.085]{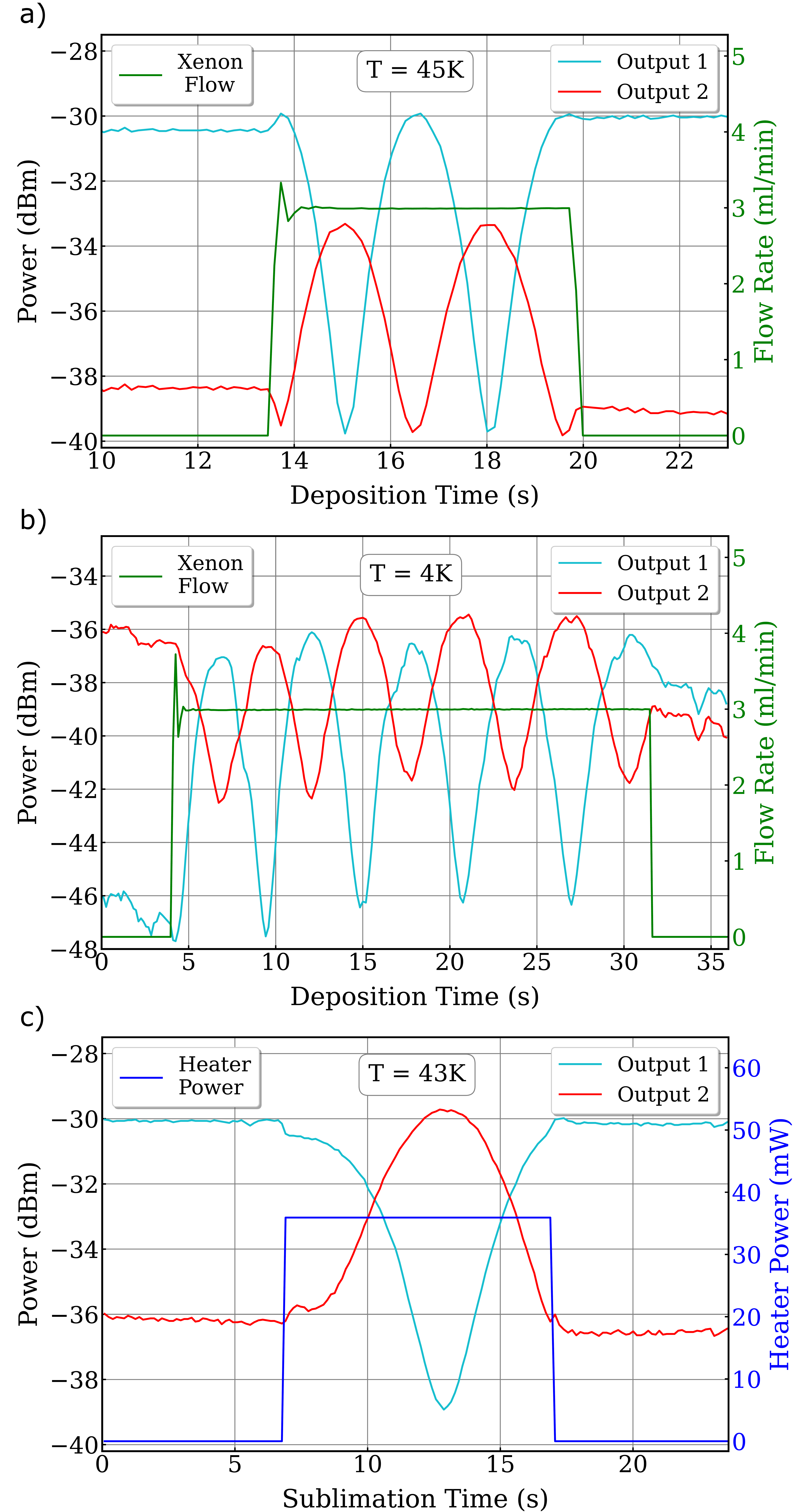}
    \caption{Change in the output of an integrated MZI with Xe deposition and sublimation. \newline Measured optical power output at each arm of the MZI, showing: a) $ \sim4~\pi$ phase shift as 0.3~$ml$ of Xe is introduced into the chamber at a rate of 3~$ml/min$. The total cumulative Xe volume in the vessel before this measurement was 1.5~$ml$ at a temperature of 45~$K$. b) $\sim9\pi$ phase shift as 1.45~$ml$ of Xe is introduced into the chamber at a rate of 3 $ml/min$. Xe thickness before this deposition is estimated to be 30~$nm$ at a temperature of 4~$K$. c) $\sim2~\pi$ phase shift as 37~$mW$ of power is applied for 10 $s$ to the on-chip heaters. This data corresponds to step 8 in testing the sublimation of thin Xe films at a temperature of 43~$K$ (Fig.~\ref{fig:Sublimation}d).}
    \label{fig:Phase_Shift}
\end{figure}

Figure \ref{fig:Phase_Shift}a shows the optical power output with time, alongside the heater power applied to obtain a $\Delta\phi$ of $\sim 2\pi$, taken from thin-film testing. We see how desorption occurs over the entire duration of the heater on-state, with the optical power balance remaining reasonably stable while no heater power is applied. Figure \ref{fig:Phase_Shift}b shows, comparatively, the response of the MZI's optical output to the flow of Xe over the circuit, resulting in a $\Delta\phi$ of $\sim4\pi$. Figure \ref{fig:Phase_Shift}c shows the same behavior at SNSPD-friendly temperatures of 4~$K$. In both cases, the film grows steadily as long as the Xe flow rate remains stable, with no significant changes in the absence of Xe flow.

\section{Discussion}

Our results demonstrate the feasibility of re-configuring and locally tuning individual components in PICs by means of cladding layers of frozen gas. 
We chose Xe due to its stability in a convenient temperature and pressure range \cite{honig1960vapor}, which lends itself to more reproducible results (\cite{Mosor2005ScanningXenon}, Fig.~\ref{fig:Ng_with_Vol}). 
Similar approaches can be implemented for different environmental conditions and different substances, suggesting the potential for room-temperature operation.

More immediate benefits of Xe come from the minimal refractive index difference between SiO$_{2}$ ($\eta = $1.44) and solid Xe ($\eta = $1.47 as found in \cite{GRACE2017204}). In particular, this benefits the high-index contrast attainable with Si waveguides and helps reduce insertion loss for a device of this kind. Photonic platforms based on core materials with lower refractive index, such as aluminium nitride (AlN) or silicon nitride (Si$_{3}$N$_{4}$), are likely to perform similarly, or even better due to their larger effective mode area.

The CLM device we present here exhibits under-etching of the areas where the cladding was removed, as evidenced by discrepancies between the simulated and experimentally observed group indices as highlighted in the simulation results. Our model is accurate enough to infer from collected data the under-etch depth to be 200$nm$. Ideal devices would have zero under-etch to minimize mode-mismatch losses. Inevitably, this would be supplemented by a slightly larger device footprint to compensate for cladding that would otherwise be missing underneath the waveguide. Under-etch implemented intentionally could instead reduce device footprint, exploiting the trade-off between mode-mismatch losses and propagation losses.
CLM-ready devices introduce an element of fragility to the circuit due to an inherent coupling of the circuit to the local environment. Consequently, cleanliness and environmental isolation become fundamental factors in their storage and operation. Cryogenic environments are in a vacuum as a standard, which leaves storage alone as a criticality when compared to fully clad circuits.

The precision to which we can set a certain phase shift comes from our ability to dial-in a certain effective index. We target a specific film thickness either on the deposition of Xe or on its sublimation. 
A linear sublimation response translates into a nearly-exponential phase response of the MZI, as seen in Fig.~\ref{fig:Ng_with_Vol}. We observe three distinct trends in this relationship, which we define as thin ($<60$~nm), thick and saturated films ($>600$~nm).
Figure ~\ref{fig:Sublimation}(a-b) infers specific sublimation rates. For deposition, we fit early data for thin films and estimate deposition rates of between 1.6 and 1.06~$nm/s$. Note how such rates are partially limited in our setup by the v-groove array (VGA), covering the unclad spiral (see Fig.~\ref{fig:Rendering}g \& c respectively).
For example, depositing Xe until the film saturates on a 70~$\mu m$-long device will typically yield a total phase change of 0.009~$\pi / s$, as estimated from lowest deposition rates. The accumulated phase drops to 0.006~$\pi / s$, if a thin film was already present at the start of the deposition. The precision of CLM varies over a deposition cycle, becoming exponentially more precise as the layer grows, in correspondence with a reduced effect of the thickness on the phase change.

Sublimation rates of Xe decrease rapidly away from the sublimation temperature as described by Fig.~\ref{fig:Rendering}a, suggesting that for localized heat delivery, neighboring Xe will experience little to no change in thickness. This is evidenced by the lack of observed sublimation at $P_{Total}$ of 18.8 $mW$ and lower powers. This power threshold is strongly dependent on the geometry of the device (as well as the cooling power of the cryostat) and is expected to be much lower for smaller, more compact structures.

In a similar fashion to the $V_{\pi}$ metric used for optical modulators \cite{jacques2019optimization}, we define the $L_{\pi}$ as the minimum exposed length of waveguide in an MZI arm necessary to obtain a $\pi$ phase shift. Consequently, for the extreme case of saturated films, $L_{\pi}$ is 12.3$\pm$0.3~$\mu m$, while a more conservative thin film case (23.9$\pm$7.7~$nm$ thickness), the obtained $L_{\pi}$ is 42.5$\pm$0.1~$\mu m$. Coupled with the TOPS comparison in the sublimation results, the range of lengths observed here suggests that the exposed waveguide sections in a CLM device can be significantly shorter than typical thermo-optic \cite{jacques2019optimization} or electro-optic phase-shifting devices \cite{eltes2020integrated}.

The most phase-stable operating conditions for a phase-shifting device lie between thin films and saturated films (for environmental conditions similar to this experiment), suggesting a minimum exposed waveguide length of 70~$\mu m$ for at least 4 $\pi$ of tunability. Estimates for the device size will vary for different photonic platforms, in line with the index contrast between core and cladding materials.
Our results show that we can deposit Xe at 45~$K$, tune the layer thickness, and then work at ideal SNSPD temperatures (4~$K$).
We have also demonstrated the viability of Xe deposition directly at the base 4~$K$ temperature, suggesting that CLM can be performed entirely at base-temperature, whilst minimizing non-Xe contamination.

CLM relies on the physical deposition of Xe and its subsequent sublimation, an intrinsically slow process. Its speed is not comparable to conventional switching mechanisms in integrated photonics. 
Despite this, CLM can still be used effectively for a systematic and sequential pre-calibration of the PIC's configurable components, without a meaningful impact on the power budget. Therefore it is possible to deploy critically larger reconfigurable circuits (including photon-sources, filters, etc.) without any additional power overhead. Within these circuits it is also worth noting that such large tunability suggests the encouraging use-case of CLM as a fabrication tolerant, on-chip tunable optical delay line.

Our results provide a substantial contribution to the quest towards a fully integrated quantum photonic platform, taking advantage of the extreme temperature conditions required for the operation of superconducting devices. Specifically, our approach facilitates the use of on-chip SNSPDs with photonic integrated circuits embedding reconfigurable components that draw no power once configured, freeing up the entire power budget for the controllable components that are required to process quantum information.
We envisage how the scheme we have developed offers opportunities for further exploration: for example, direct-laser-writing (DLW, \cite{davis1996writing,gehring2020reconfigurable}) could be used directly to tune the Xe film with zero cross-talk and no need for on-chip electrical connections. 

Moreover, CLM is not restricted to switches and is intrinsically compatible with any device that uses effective/group indices to determine performance characteristics, with devices more sensitive to fabrication tolerances benefiting the most. MMIs are known for their fabrication tolerance, but directional couplers, sub-wavelength gratings, and grating couplers are all examples that could be optimized in-situ using our approach.

\paragraph*{Funding.}
B.M.B. was supported by the Quantum Engineering Centre for Doctoral Training at Bristol, EPSRC Grant No. EP/L015730/1. The authors acknowledge support from the Engineering and
Physical Sciences Research Council.

\paragraph*{Acknowledgments.}
We would like to thank Andy Murray for assistance with the sample preparation. J.B. wants to thank Jake Kennard, Josh W. Silverstone and Alberto Santamato for useful discussions.

\paragraph*{Disclosures.}
J.B, G.E.V-G, and B.M.B declare UK patent application number 2009590.7 on the apparatus and method described in the paper.

\subsection*{Author Contributions}
J.B conceived the idea, A.A.G designed the experiment, P.J post-processed the photonic circuit. B.M.B and G.E.V-G assembled the experimental apparatus and ran simulations. B.M.B designed the control software, performed the experiment, and analyzed the data. J.B supervised the project. B.M.B. wrote the manuscript with the support from all the authors.

\paragraph*{Data Availability.}
All data needed to evaluate the conclusions in the paper are present in the paper and/or the Supplementary Materials. The primary data that supports the plots and other findings reported in this study can be requested from the corresponding author.

\paragraph*{Supplemental Document.} See the supplementary materials for supporting content.


\bibliographystyle{naturemag}

\begin{thebibliography}{9}

\expandafter\ifx\csname url\endcsname\relax
  \def\url#1{\texttt{#1}}\fi
\expandafter\ifx\csname urlprefix\endcsname\relax\def\urlprefix{URL }\fi
\providecommand{\bibinfo}[2]{#2}
\providecommand{\eprint}[2][]{\url{#2}}

\bibitem{o2009photonic}
\bibinfo{author}{O'Brien, J.~L.}, \bibinfo{author}{Furusawa, A.} \&
  \bibinfo{author}{Vu{\v{c}}kovi{\'c}, J.}
\newblock \bibinfo{title}{"photonic quantum technologies"}.
\newblock \emph{\bibinfo{journal}{Nature Photonics}}
  \textbf{\bibinfo{volume}{3}}, \bibinfo{pages}{687} (\bibinfo{year}{2009}).

\bibitem{qiang2018large}
\bibinfo{author}{Qiang, X.} \emph{et~al.}
\newblock \bibinfo{title}{Large-scale silicon quantum photonics implementing
  arbitrary two-qubit processing}.
\newblock \emph{\bibinfo{journal}{Nature photonics}}
  \textbf{\bibinfo{volume}{12}}, \bibinfo{pages}{534--539}
  (\bibinfo{year}{2018}).

\bibitem{Obrien2007opticalqc}
\bibinfo{author}{O'Brien, J.~L.}
\newblock \bibinfo{title}{{Optical quantum computing}}.
\newblock \emph{\bibinfo{journal}{Science}} \textbf{\bibinfo{volume}{318}},
  \bibinfo{pages}{390--397} (\bibinfo{year}{2007}).

\bibitem{silverstone2015qubit}
\bibinfo{author}{Silverstone, J.~W.} \emph{et~al.}
\newblock \bibinfo{title}{Qubit entanglement between ring-resonator photon-pair
  sources on a silicon chip}.
\newblock \emph{\bibinfo{journal}{Nature communications}}
  \textbf{\bibinfo{volume}{6}}, \bibinfo{pages}{1--7} (\bibinfo{year}{2015}).

\bibitem{wang2019integrated}
\bibinfo{author}{Wang, J.}, \bibinfo{author}{Sciarrino, F.},
  \bibinfo{author}{Laing, A.} \& \bibinfo{author}{Thompson, M.~G.}
\newblock \bibinfo{title}{Integrated photonic quantum technologies}.
\newblock \emph{\bibinfo{journal}{Nature Photonics}} \bibinfo{pages}{1--12}
  (\bibinfo{year}{2019}).

\bibitem{silverstone2016}
\bibinfo{author}{Silverstone, J.}, \bibinfo{author}{Bonneau, D.},
  \bibinfo{author}{O'Brien, J.~L.} \& \bibinfo{author}{Thompson, M.~G.}
\newblock \bibinfo{title}{Silicon quantum photonics}.
\newblock \emph{\bibinfo{journal}{IEEE}} \textbf{\bibinfo{volume}{22}},
  \bibinfo{pages}{6700113} (\bibinfo{year}{2016}).

\bibitem{doi:10.1063/1.3657518}
\bibinfo{author}{Sprengers, J.~P.} \emph{et~al.}
\newblock \bibinfo{title}{{Waveguide superconducting single-photon detectors
  for integrated quantum photonic circuits}}.
\newblock \emph{\bibinfo{journal}{Applied Physics Letters}}
  \textbf{\bibinfo{volume}{99}}, \bibinfo{pages}{181110}
  (\bibinfo{year}{2011}).

\bibitem{Reddy:19}
\bibinfo{author}{Reddy, D.~V.} \emph{et~al.}
\newblock \bibinfo{title}{{Exceeding 95{\%} system efficiency within the
  telecom C-band in superconducting nanowire single photon detectors}}.
\newblock In \emph{\bibinfo{booktitle}{Conference on Lasers and
  Electro-Optics}}, \bibinfo{pages}{FF1A.3} (\bibinfo{publisher}{Optical
  Society of America}, \bibinfo{year}{2019}).

\bibitem{Pernice_2012}
\bibinfo{author}{Pernice, W. H.~P.} \emph{et~al.}
\newblock \bibinfo{title}{{High-speed and high-efficiency travelling wave
  single-photon detectors embedded in nanophotonic circuits}}.
\newblock \emph{\bibinfo{journal}{Nature Communications}}
  \textbf{\bibinfo{volume}{3}} (\bibinfo{year}{2012}).

\bibitem{tyler2016modelling}
\bibinfo{author}{Tyler, N.~A.} \emph{et~al.}
\newblock \bibinfo{title}{Modelling superconducting nanowire single photon
  detectors in a waveguide cavity}.
\newblock \emph{\bibinfo{journal}{Optics express}}
  \textbf{\bibinfo{volume}{24}}, \bibinfo{pages}{8797--8808}
  (\bibinfo{year}{2016}).

\bibitem{reithmaier2013chip}
\bibinfo{author}{Reithmaier, G.} \emph{et~al.}
\newblock \bibinfo{title}{On-chip time resolved detection of quantum dot
  emission using integrated superconducting single photon detectors}.
\newblock \emph{\bibinfo{journal}{Scientific reports}}
  \textbf{\bibinfo{volume}{3}}, \bibinfo{pages}{1901} (\bibinfo{year}{2013}).

\bibitem{sahin2013waveguide}
\bibinfo{author}{Sahin, D.} \emph{et~al.}
\newblock \bibinfo{title}{Waveguide photon-number-resolving detectors for
  quantum photonic integrated circuits}.
\newblock \emph{\bibinfo{journal}{Applied Physics Letters}}
  \textbf{\bibinfo{volume}{103}}, \bibinfo{pages}{111116}
  (\bibinfo{year}{2013}).

\bibitem{schuck2016quantum}
\bibinfo{author}{Schuck, C.} \emph{et~al.}
\newblock \bibinfo{title}{Quantum interference in heterogeneous
  superconducting-photonic circuits on a silicon chip}.
\newblock \emph{\bibinfo{journal}{Nature communications}}
  \textbf{\bibinfo{volume}{7}}, \bibinfo{pages}{1--7} (\bibinfo{year}{2016}).

\bibitem{schuck2013waveguide}
\bibinfo{author}{Schuck, C.}, \bibinfo{author}{Pernice, W.~H.} \&
  \bibinfo{author}{Tang, H.~X.}
\newblock \bibinfo{title}{Waveguide integrated low noise nbtin nanowire
  single-photon detectors with milli-hz dark count rate}.
\newblock \emph{\bibinfo{journal}{Scientific reports}}
  \textbf{\bibinfo{volume}{3}}, \bibinfo{pages}{1893} (\bibinfo{year}{2013}).

\bibitem{najafi2015chip}
\bibinfo{author}{Najafi, F.} \emph{et~al.}
\newblock \bibinfo{title}{On-chip detection of non-classical light by scalable
  integration of single-photon detectors}.
\newblock \emph{\bibinfo{journal}{Nature communications}}
  \textbf{\bibinfo{volume}{6}}, \bibinfo{pages}{1--8} (\bibinfo{year}{2015}).

\bibitem{Bogaerts2020}
\bibinfo{author}{Bogaerts, W.} \emph{et~al.}
\newblock \bibinfo{title}{Programmable photonic circuits}.
\newblock \emph{\bibinfo{journal}{Nature}} \textbf{\bibinfo{volume}{586}},
  \bibinfo{pages}{207--216} (\bibinfo{year}{2020}).

\bibitem{Watts:13}
\bibinfo{author}{Watts, M.~R.} \emph{et~al.}
\newblock \bibinfo{title}{{Adiabatic thermo-optic Mach-Zehnder switch}}.
\newblock \emph{\bibinfo{journal}{Opt. Lett.}} \textbf{\bibinfo{volume}{38}},
  \bibinfo{pages}{733--735} (\bibinfo{year}{2013}).

\bibitem{silverstone2014chip}
\bibinfo{author}{Silverstone, J.~W.} \emph{et~al.}
\newblock \bibinfo{title}{{On-chip quantum interference between silicon
  photon-pair sources}}.
\newblock \emph{\bibinfo{journal}{Nature Photonics}}
  \textbf{\bibinfo{volume}{8}}, \bibinfo{pages}{104} (\bibinfo{year}{2014}).

\bibitem{8613782}
\bibinfo{author}{Eltes, F.} \emph{et~al.}
\newblock \bibinfo{title}{{A BaTiO3-Based Electro-Optic Pockels Modulator
  Monolithically Integrated on an Advanced Silicon Photonics Platform}}.
\newblock \emph{\bibinfo{journal}{Journal of Lightwave Technology}}
  \textbf{\bibinfo{volume}{37}}, \bibinfo{pages}{1456--1462}
  (\bibinfo{year}{2019}).

\bibitem{Reed2010}
\bibinfo{author}{Reed, G.~T.}, \bibinfo{author}{Mashanovich, G.},
  \bibinfo{author}{Gardes, F.~Y.} \& \bibinfo{author}{Thomson, D.~J.}
\newblock \bibinfo{title}{{Silicon optical modulators}}.
\newblock \emph{\bibinfo{journal}{Nature Photonics}}
  \textbf{\bibinfo{volume}{4}}, \bibinfo{pages}{518--526}
  (\bibinfo{year}{2010}).

\bibitem{wang2017experimental}
\bibinfo{author}{Wang, J.} \emph{et~al.}
\newblock \bibinfo{title}{Experimental quantum hamiltonian learning}.
\newblock \emph{\bibinfo{journal}{Nature Physics}}
  \textbf{\bibinfo{volume}{13}}, \bibinfo{pages}{551--555}
  (\bibinfo{year}{2017}).

\bibitem{Faruque:18}
\bibinfo{author}{Faruque, I.~I.}, \bibinfo{author}{Sinclair, G.~F.},
  \bibinfo{author}{Bonneau, D.}, \bibinfo{author}{Rarity, J.~G.} \&
  \bibinfo{author}{Thompson, M.~G.}
\newblock \bibinfo{title}{{On-chip quantum interference with heralded photons
  from two independent micro-ring resonator sources in silicon photonics}}.
\newblock \emph{\bibinfo{journal}{Opt. Express}} \textbf{\bibinfo{volume}{26}},
  \bibinfo{pages}{20379--20395} (\bibinfo{year}{2018}).

\bibitem{llewellyn2020chip}
\bibinfo{author}{Llewellyn, D.} \emph{et~al.}
\newblock \bibinfo{title}{Chip-to-chip quantum teleportation and multi-photon
  entanglement in silicon}.
\newblock \emph{\bibinfo{journal}{Nature Physics}}
  \textbf{\bibinfo{volume}{16}}, \bibinfo{pages}{148--153}
  (\bibinfo{year}{2020}).

\bibitem{7463458}
\bibinfo{author}{Elshaari, A.~W.}, \bibinfo{author}{Zadeh, I.~E.},
  \bibinfo{author}{J{\"{o}}ns, K.~D.} \& \bibinfo{author}{Zwiller, V.}
\newblock \bibinfo{title}{{Thermo-Optic Characterization of Silicon Nitride
  Resonators for Cryogenic Photonic Circuits}}.
\newblock \emph{\bibinfo{journal}{IEEE Photonics Journal}}
  \textbf{\bibinfo{volume}{8}}, \bibinfo{pages}{1--9} (\bibinfo{year}{2016}).

\bibitem{eltes2020integrated}
\bibinfo{author}{Eltes, F.} \emph{et~al.}
\newblock \bibinfo{title}{An integrated optical modulator operating at
  cryogenic temperatures}.
\newblock \emph{\bibinfo{journal}{Nature Materials}} \bibinfo{pages}{1--5}
  (\bibinfo{year}{2020}).

\bibitem{Mosor2005ScanningXenon}
\bibinfo{author}{Mosor, S.} \emph{et~al.}
\newblock \bibinfo{title}{{Scanning a photonic crystal slab nanocavity by
  condensation of xenon}}.
\newblock \emph{\bibinfo{journal}{Applied Physics Letters}}
  \textbf{\bibinfo{volume}{87}}, \bibinfo{pages}{141105}
  (\bibinfo{year}{2005}).

\bibitem{li2015coherent}
\bibinfo{author}{Li, L.} \emph{et~al.}
\newblock \bibinfo{title}{Coherent spin control of a nanocavity-enhanced qubit
  in diamond}.
\newblock \emph{\bibinfo{journal}{Nature communications}}
  \textbf{\bibinfo{volume}{6}}, \bibinfo{pages}{1--7} (\bibinfo{year}{2015}).

\bibitem{kaneda2019high}
\bibinfo{author}{Kaneda, F.} \& \bibinfo{author}{Kwiat, P.~G.}
\newblock \bibinfo{title}{High-efficiency single-photon generation via
  large-scale active time multiplexing}.
\newblock \emph{\bibinfo{journal}{Science advances}}
  \textbf{\bibinfo{volume}{5}}, \bibinfo{pages}{eaaw8586}
  (\bibinfo{year}{2019}).

\bibitem{articleb}
\bibinfo{author}{Bogaerts, W.} \& \bibinfo{author}{Chrostowski, L.}
\newblock \bibinfo{title}{{Silicon Photonics Circuit Design: Methods, Tools and
  Challenges}}.
\newblock \emph{\bibinfo{journal}{Laser {\&} Photonics Reviews}}
  \textbf{\bibinfo{volume}{12}}, \bibinfo{pages}{1700237}
  (\bibinfo{year}{2018}).

\bibitem{296191}
\bibinfo{author}{Besse, P.~A.}, \bibinfo{author}{Bachmann, M.},
  \bibinfo{author}{Melchior, H.}, \bibinfo{author}{Soldano, L.~B.} \&
  \bibinfo{author}{Smit, M.~K.}
\newblock \bibinfo{title}{{Optical bandwidth and fabrication tolerances of
  multimode interference couplers}}.
\newblock \emph{\bibinfo{journal}{Journal of Lightwave Technology}}
  \textbf{\bibinfo{volume}{12}}, \bibinfo{pages}{1004--1009}
  (\bibinfo{year}{1994}).

\bibitem{Rajarajan:99}
\bibinfo{author}{Rajarajan, M.}, \bibinfo{author}{Rahman, B. M.~A.} \&
  \bibinfo{author}{Grattan, K. T.~V.}
\newblock \bibinfo{title}{{A Rigorous Comparison of the Performance of
  Directional Couplers with Multimode Interference Devices}}.
\newblock \emph{\bibinfo{journal}{J. Lightwave Technol.}}
  \textbf{\bibinfo{volume}{17}}, \bibinfo{pages}{243} (\bibinfo{year}{1999}).

\bibitem{honig1960vapor}
\bibinfo{author}{Honig, R.~E.} \& \bibinfo{author}{Hook, H.~O.}
\newblock \bibinfo{title}{{Vapor pressure data for some common gases}}.
\newblock \emph{\bibinfo{journal}{RCA review}} \textbf{\bibinfo{volume}{21}},
  \bibinfo{pages}{360--368} (\bibinfo{year}{1960}).

\bibitem{fabricius1992gas}
\bibinfo{author}{Fabricius, N.}, \bibinfo{author}{Gauglitz, G.} \&
  \bibinfo{author}{Ingenhoff, J.}
\newblock \bibinfo{title}{A gas sensor based on an integrated optical
  mach-zehnder interferometer}.
\newblock \emph{\bibinfo{journal}{Sensors and Actuators B: Chemical}}
  \textbf{\bibinfo{volume}{7}}, \bibinfo{pages}{672--676}
  (\bibinfo{year}{1992}).

\bibitem{GRACE2017204}
\bibinfo{author}{Grace, E.}, \bibinfo{author}{Butcher, A.},
  \bibinfo{author}{Monroe, J.} \& \bibinfo{author}{Nikkel, J.~A.}
\newblock \bibinfo{title}{{Index of refraction, Rayleigh scattering length, and
  Sellmeier coefficients in solid and liquid argon and xenon}}.
\newblock \emph{\bibinfo{journal}{Nuclear Instruments and Methods in Physics
  Research Section A: Accelerators, Spectrometers, Detectors and Associated
  Equipment}} \textbf{\bibinfo{volume}{867}}, \bibinfo{pages}{204--208}
  (\bibinfo{year}{2017}).

\bibitem{jousten2018handbuch}
\bibinfo{author}{Jousten, K.}
\newblock \emph{\bibinfo{title}{Handbuch Vakuumtechnik}},
  chap.~\bibinfo{chapter}{7}, \bibinfo{pages}{668}
  (\bibinfo{publisher}{Springer-Verlag}, \bibinfo{year}{2018}).

\bibitem{frey2006temperature}
\bibinfo{author}{Frey, B.~J.}, \bibinfo{author}{Leviton, D.~B.} \&
  \bibinfo{author}{Madison, T.~J.}
\newblock \bibinfo{title}{Temperature-dependent refractive index of silicon and
  germanium}.
\newblock In \emph{\bibinfo{booktitle}{Optomechanical Technologies for
  Astronomy}}, vol. \bibinfo{volume}{6273}, \bibinfo{pages}{62732J}
  (\bibinfo{organization}{International Society for Optics and Photonics},
  \bibinfo{year}{2006}).

\bibitem{leviton2006temperature}
\bibinfo{author}{Leviton, D.~B.} \& \bibinfo{author}{Frey, B.~J.}
\newblock \bibinfo{title}{Temperature-dependent absolute refractive index
  measurements of synthetic fused silica}.
\newblock In \emph{\bibinfo{booktitle}{Optomechanical Technologies for
  Astronomy}}, vol. \bibinfo{volume}{6273}, \bibinfo{pages}{62732K}
  (\bibinfo{organization}{International Society for Optics and Photonics},
  \bibinfo{year}{2006}).

\bibitem{Taillaert_2006}
\bibinfo{author}{Taillaert, D.} \emph{et~al.}
\newblock \bibinfo{title}{{Grating Couplers for Coupling between Optical Fibers
  and Nanophotonic Waveguides}}.
\newblock \emph{\bibinfo{journal}{Japanese Journal of Applied Physics}}
  \textbf{\bibinfo{volume}{45}}, \bibinfo{pages}{6071--6077}
  (\bibinfo{year}{2006}).

\bibitem{doi:10.1021/cr60119a001}
\bibinfo{author}{Thomson, G.~W.}
\newblock \bibinfo{title}{{The Antoine Equation for Vapor-pressure Data.}}
\newblock \emph{\bibinfo{journal}{Chemical Reviews}}
  \textbf{\bibinfo{volume}{38}}, \bibinfo{pages}{1--39} (\bibinfo{year}{1946}).

\bibitem{lu2017performance}
\bibinfo{author}{Lu, Z.} \emph{et~al.}
\newblock \bibinfo{title}{Performance prediction for silicon photonics
  integrated circuits with layout-dependent correlated manufacturing
  variability}.
\newblock \emph{\bibinfo{journal}{Optics express}}
  \textbf{\bibinfo{volume}{25}}, \bibinfo{pages}{9712--9733}
  (\bibinfo{year}{2017}).

\bibitem{biberman2012ultralow}
\bibinfo{author}{Biberman, A.}, \bibinfo{author}{Shaw, M.~J.},
  \bibinfo{author}{Timurdogan, E.}, \bibinfo{author}{Wright, J.~B.} \&
  \bibinfo{author}{Watts, M.~R.}
\newblock \bibinfo{title}{Ultralow-loss silicon ring resonators}.
\newblock \emph{\bibinfo{journal}{Optics letters}}
  \textbf{\bibinfo{volume}{37}}, \bibinfo{pages}{4236--4238}
  (\bibinfo{year}{2012}).

\bibitem{jacques2019optimization}
\bibinfo{author}{Jacques, M.} \emph{et~al.}
\newblock \bibinfo{title}{{Optimization of thermo-optic phase-shifter design
  and mitigation of thermal crosstalk on the SOI platform}}.
\newblock \emph{\bibinfo{journal}{Optics express}}
  \textbf{\bibinfo{volume}{27}}, \bibinfo{pages}{10456--10471}
  (\bibinfo{year}{2019}).

\bibitem{davis1996writing}
\bibinfo{author}{Davis, K.~M.}, \bibinfo{author}{Miura, K.},
  \bibinfo{author}{Sugimoto, N.} \& \bibinfo{author}{Hirao, K.}
\newblock \bibinfo{title}{Writing waveguides in glass with a femtosecond
  laser}.
\newblock \emph{\bibinfo{journal}{Optics letters}}
  \textbf{\bibinfo{volume}{21}}, \bibinfo{pages}{1729--1731}
  (\bibinfo{year}{1996}).
  
\bibitem{gehring2020reconfigurable}
\bibinfo{author}{Gehring, H.}, \bibinfo{author}{Blaicher, M.},
  \bibinfo{author}{Grottke, T.} \& \bibinfo{author}{Pernice, W.~H.}
\newblock \bibinfo{title}{Reconfigurable nanophotonic circuitry enabled by
  direct-laser-writing}.
\newblock \emph{\bibinfo{journal}{IEEE Journal of Selected Topics in Quantum
  Electronics}} \textbf{\bibinfo{volume}{26}}, \bibinfo{pages}{1--5}
  (\bibinfo{year}{2020}).

\bibitem{Eltes2018bto}
\bibinfo{author}{Eltes, F.} \emph{et~al.}
\newblock \bibinfo{title}{{First cryogenic electro-optic switch on silicon with
  high bandwidth and low power tunability}}.
\newblock In \emph{\bibinfo{booktitle}{2018 IEEE International Electron Devices
  Meeting (IEDM)}}, \bibinfo{pages}{23.1.1--23.1.4} (\bibinfo{publisher}{IEEE},
  \bibinfo{year}{2018}).

\bibitem{Liu2016_optproc}
\bibinfo{author}{Liu, W.} \emph{et~al.}
\newblock \bibinfo{title}{{A fully reconfigurable photonic integrated signal
  processor}}.
\newblock \emph{\bibinfo{journal}{Nature Photonics}}
  \textbf{\bibinfo{volume}{10}}, \bibinfo{pages}{190--195}
  (\bibinfo{year}{2016}).

\bibitem{Harris2018lin_photoproc}
\bibinfo{author}{Harris, N.~C.} \emph{et~al.}
\newblock \bibinfo{title}{{Linear programmable nanophotonic processors}}.
\newblock \emph{\bibinfo{journal}{Optica}} \textbf{\bibinfo{volume}{5}},
  \bibinfo{pages}{1623} (\bibinfo{year}{2018}).

\bibitem{Komma2012thermo_optic}
\bibinfo{author}{Komma, J.}, \bibinfo{author}{Schwarz, C.},
  \bibinfo{author}{Hofmann, G.}, \bibinfo{author}{Heinert, D.} \&
  \bibinfo{author}{Nawrodt, R.}
\newblock \bibinfo{title}{{Thermo-optic coefficient of silicon at 1550 nm and
  cryogenic temperatures}}.
\newblock \emph{\bibinfo{journal}{Applied Physics Letters}}
  \textbf{\bibinfo{volume}{101}}, \bibinfo{pages}{041905}
  (\bibinfo{year}{2012}).

\bibitem{Villareal2018thesis}
\bibinfo{author}{Villareal-Garcia, G.}
\newblock \emph{\bibinfo{title}{{Thermo--optic phase shifters for integrated
  photonics at low temperatures}}}.
\newblock Ph.D. thesis, \bibinfo{school}{University of Bristol}
  (\bibinfo{year}{2018}).

\bibitem{chrostowski2015photonics}
\bibinfo{author}{Chrostowski, L.} \& \bibinfo{author}{Hochberg, M.}
\newblock \emph{\bibinfo{title}{Silicon photonics design: from devices to
  systems}} (\bibinfo{publisher}{Cambridge University Press},
  \bibinfo{year}{2015}).

\bibitem{agrawal2004lightwave}
\bibinfo{author}{Agrawal, G.~P.}
\newblock \emph{\bibinfo{title}{Lightwave technology: components and devices}},
  vol.~\bibinfo{volume}{1} (\bibinfo{publisher}{John Wiley \& Sons},
  \bibinfo{year}{2004}).

\bibitem{Bruck2016allopticalmod}
\bibinfo{author}{Bruck, R.} \emph{et~al.}
\newblock \bibinfo{title}{{All-optical spatial light modulator for
  reconfigurable silicon photonic circuits}}.
\newblock \emph{\bibinfo{journal}{Optica}} \textbf{\bibinfo{volume}{3}},
  \bibinfo{pages}{396} (\bibinfo{year}{2016}).
\newblock \eprint{1601.06679}.

\bibitem{Timurdogan2017pinmodulat}
\bibinfo{author}{Timurdogan, E.}, \bibinfo{author}{Poulton, C.~V.},
  \bibinfo{author}{Byrd, M.~J.} \& \bibinfo{author}{Watts, M.~R.}
\newblock \bibinfo{title}{{Electric field-induced second-order nonlinear
  optical effects in silicon waveguides}}.
\newblock \emph{\bibinfo{journal}{Nature Photonics}}
  \textbf{\bibinfo{volume}{11}}, \bibinfo{pages}{200--206}
  (\bibinfo{year}{2017}).

\bibitem{BERMAN1996327}
\bibinfo{author}{Berman, A.}
\newblock \bibinfo{title}{{Water vapor in vacuum systems}}.
\newblock \emph{\bibinfo{journal}{Vacuum}} \textbf{\bibinfo{volume}{47}},
  \bibinfo{pages}{327--332} (\bibinfo{year}{1996}).

\end{thebibliography}

\clearpage

\begin{centering}
{\LARGE\bfseries Zero-power calibration of photonic circuits at cryogenic temperatures: supplemental document}

\section*{This PDF file includes:}
Supplementary Text\\
Figs. S1 to S5\\
References (45-54)

\end{centering}

\section*{Low--power reconfigurability for cryogenic photonics}

\normalsize{T}he practicality of the quantum devices employed in delivering successful protocols has been mostly based upon designing and testing noise--resilient approaches in the data processing. 
In this way, it was possible to avoid complex quantum circuits with width and depth too high to be handled within current photonic technological limits, or the demand for a cryogenic environment when using solid--state quantum sensors.

However, software developments alone are likely not enough to push the application boundaries of quantum technologies in many cases of interest.
Here, we picked a specific technological challenge: the reconfigurability of photonic devices in a cryogenic environment.
The ability to reconfigure a photonic circuit is key in deploying a scalable photonic architecture for LOQC, as well as implementing digital quantum simulations. 

Thermo-optic phase shifters (TOPS) are limited to operate in the range of $MHz$ \cite{silverstone2016,Eltes2018bto}. This can be a severe limitation in several applications. In LOQC, the feed-forward scheme requires high--speed reconfiguration, to avoid incurring  unacceptable losses when employing on--purpose delay lines\cite{Obrien2007opticalqc,silverstone2016}. 
High--speed reconfigurability is also a necessary requirement for classical photonic processors, recently pursued as a promising route towards achieving lower power and higher speed data processing, when compared to their electronic counterparts (e.g. \cite{Liu2016_optproc,Harris2018lin_photoproc}).

A perhaps even stronger limitation appears when considering cryogenic operation scenarios, required either by efficient superconducting classical processors \cite{Eltes2018bto}, or by the monolithic integration of SNSPDs in the quantum photonics realm.

It is known that the TO coefficient rapidly drops below 10$^{-5}K^{-1}$ at temperatures below 50$K$ \cite{Komma2012thermo_optic}. This might be (at least partially) compensated by the corresponding decrease in the specific heat coefficient \cite{Villareal2018thesis}, however, even if allowing for room--temperature power consumption, TO heaters are no scalable approach in a cryogenic environment. 

Additional challenges arise when considering SOI quantum photonics, as for a satisfactory modulation several performances should be targeted simultaneously:
\begin{enumerate}
    \item CMOS--compatibility, or compatibility with industrial--scale fabrication technologies for integrated photonic devices;
    \item device footprint;
    \item losses and noise introduced by the modulator;
    \item power consumption;
    \item speed / bandwidth;
    \item low temperature (1-4{$K$}) operability.
\end{enumerate}

Standard modulators in silicon photonics leverage upon the \textit{plasma dispersion effect}: altering the concentration\footnote{
Either via carrier \textit{accumulation}, \textit{injection} or \textit{depletion} \cite{Reed2010}.
} of free charges in Si it is possible to modulate the material refractive index $n$ \cite{chrostowski2015photonics,Reed2010}. Unfortunately, already at room temperature these modulators are known to introduce unwanted additional noise and losses in the device \cite{Reed2010,silverstone2016}. This drawback is made worse when moving to lower operating temperatures, because the carrier freeze--out leads to higher concentrations of dopants in the Si WG, which in turns lead to higher losses and lower bandwidth \cite{eltes2020integrated}.
All--optical modulation, if a promising candidate for classical photonic devices \cite{agrawal2004lightwave,Bruck2016allopticalmod}, is impractical in the realm of quantum photonics, where filtering out on--demand signals from an additional bright pump is possible in principle, but technologically difficult.

The failure of $\chi^{(3)}$--based approaches to deliver a suitable modulation of $n$ in Si has pushed for solutions that introduce a $\chi^{(2)}$ effect in SOI platforms, which can be achieved in several different ways.
As $\chi^{(2)}=0$ is due to the centro--symmetric structure of Si crystals, a possibility is to break such symmetry, by either strain or strong electrical fields \cite{silverstone2016}.   
If the CMOS compatibility of the first approach is yet to be proven, such an implementation has already been demonstrated with satisfactory bandwidth, limited to few $GHz$ \cite{Timurdogan2017pinmodulat}; however, the latter still requires improvements, as it introduces additional losses as high as $\sim$ 0.2 $dB$ per modulator.

Finally, a more radical and direct approach is to hybridize the SOI platform by introducing materials that natively exhibit strong non--linearities (e.g. they have $\chi^{(2)} \neq 0$) \cite{silverstone2016}. 
To achieve this, it is not necessary to completely replace the material of the WG. The confinement of the guided optical mode within the WG is high, but not perfect. Therefore, by depositing other materials in the proximity of the Silicon WG, where the evanescent field is not yet negligible, modulation of the effective refractive index is still possible. This is the approach pursued in this work. 
A natural choice for such a hybrid material was offered by LiNbO$_3$. However, even if already widely adopted in the field of classical communication, its widespread adoption has been hindered by the lack of a credible CMOS--compatible process to integrate it in SOI.

\section*{Non--volatile, cryo--compatible light manipulation via Xe condensation}
\label{snowmotivation}

Here we outline an approach to low--power reconfigurable on--chip switches, that is based on the principle of interacting with the evanescent field in the proximity of a WG, that we dub as \textit{non--volatile}.
Such naming is due to its fundamental distinction, when compared with all other technological solutions listed until now: non--volatile modulators dissipate power only when their configuration is \textit{changed}, but not when it is \textit{maintained}. 
Indeed, this is not the case for thermo-optic nor plasma-dispersion modulators: the only configurations that they can maintain without dissipating energy are the ones in absence of any external (electrical) power supply. 
However, any reconfiguration implies operating with an applied bias voltage $V \neq 0$. As soon as the external power supply is removed, such reconfiguration ceases: in our wording, it is \textit{volatile}. 

Such strategy is inspired by observing how for some experiments, most phases implemented in the circuit are held constant throughout several measurements, and sometimes even for the whole experiment. This leads to a passive level of power consumption, in line with the heat output of the thermo-optic phase shifters.
In such cases, the target is to achieve low (virtually none) power consumption, rather than fast reconfigurability. 

Here, we propose and perform a preliminary investigation of an unconventional solution: the controlled condensation of gases in the proximity of the WGs. This idea leverages upon the very challenge to power dissipation in integrated chips: their installation in a cryogenic setup, whereby the gas condensation can be obtained for free. 
The evanescent field at the surface of the cladding layer is too weak to envisage any substantial effect of replacing the SiO$_2$-vacuum interface(in a cryogenic scenario) with a SiO$_2$-X interface, with any substance X; gaseous at room temperature, but solid at the cryostat base temperature. 
Therefore, as schematically described in Fig. 1a, cladding has to be removed in order to expose a section of the WGs (a \textit{window}). 

Condensing gases on top of exposed light-confinement nanostructures is a technique already known in the literature, but its application was limited so far to the tuning of peaks in the spectral response of photonic crystal cavities \cite{Mosor2005ScanningXenon}.
More broadly, the idea to expose sections of a waveguide to the environment, \textit{transducing} changes occurring in its composition into a change in $n_{\textrm{eff}}$, was employed for sensing purposes \cite{fabricius1992gas}. 

\section*{Experimental considerations and stability analysis}

This experiment aimed to demonstrate the principle of using gas deposition as a means to fine-tune the phase of a guided optical mode, in an approach we are calling cladding layer manipulation (CLM). Beyond this it becomes important to investigate the finer details of our methodology. Here we address; 
\begin{enumerate}
    \item Geometric considerations of Xe flow for the prediction of deposition behaviour (Simulation, Fig. S1).
    \item Feasibility of dropping to 4$K$ after deposition at 40$K$ (Fig. S2).
    \item Trend in MZI stability as temperature decreases (Fig. S3).
    \item Stability of Xe film thickness over time in a vacuum of 1$\times10^{-6}$ $mBar$ (Fig. S4).
    \item Geometric considerations of a thermo-optic solution for the tuning of film thickness (Fig. S5).
\end{enumerate}
\clearpage 
\begin{figure}[h]
    \centering
    \includegraphics[width=\linewidth]{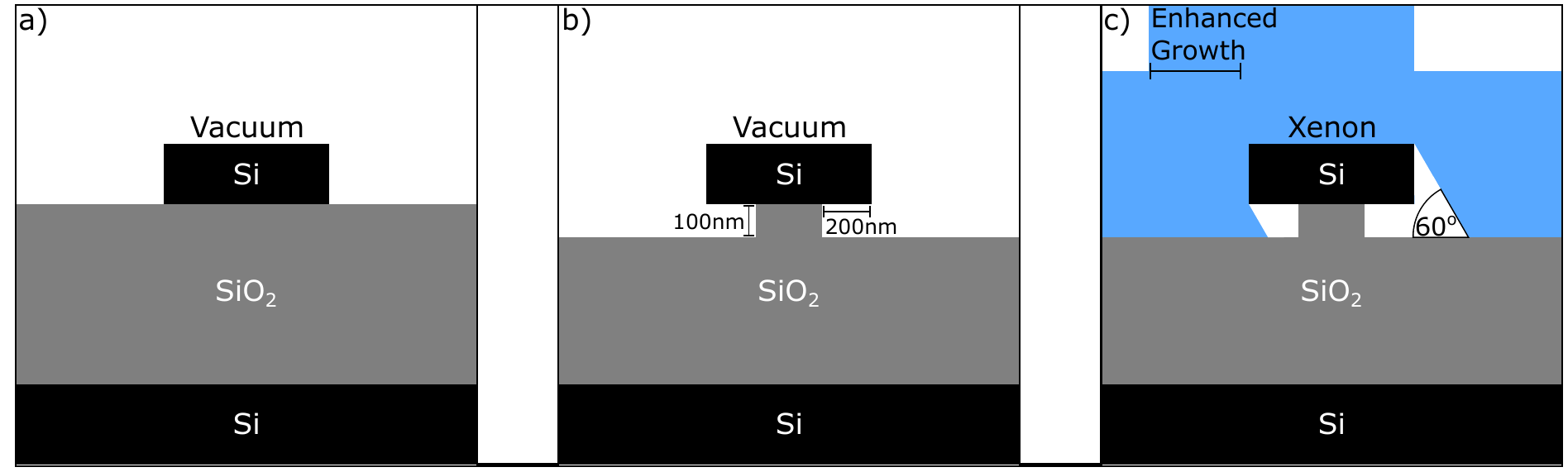}
    \caption{Diagram a) depicts the ideal non-undercut geometry of the waveguide. Diagram b) shows the undercut waveguide geometry that we used to approximately model the growth of Xe on the waveguide surface. Diagram c) demonstrates our use of asymmetric growth behavior to better represent the flow of Xe from our experimental setup.}
    \label{fig:Simulation_Geometry}
\end{figure}

We interpreted discrepancies between simulations of Xe film growth and experimental data in terms of Xe film uniformity. Any initial differences in $n_{g}$ were attributed to the waveguide geometry (Fig. S1a-b).  We simulated shadows (and associated regions of enhanced growth) to more accurately represent the device's geometry and better approximate film growth behavior (Fig. S1c). The shadow angle of 60$^\circ$ was estimated from the experiment and re-adjusted through repeated simulations; finally resulting in the Xe deposition behavior presented in Fig.~2.\\

We investigated our second point with a film of approximately 70~$nm$. After depositing the initial layers of Xe at 40~$K$ and measuring the output spectrum of the MZI, we dropped the sample to a temperature of 4~$K$ and again measured the output spectrum of the MZI (Fig. S2). We observed a phase shift per $\mu m$ of 6$\pm$1$\times$10$^{-4}~\pi$, which correlates to a phase shift of 4.2$\pm$0.7$\times$10$^{-2}~\pi$ for a 70~$\mu m$ device.\\

\begin{figure}[h]
    \centering
    \includegraphics[scale = 0.6]{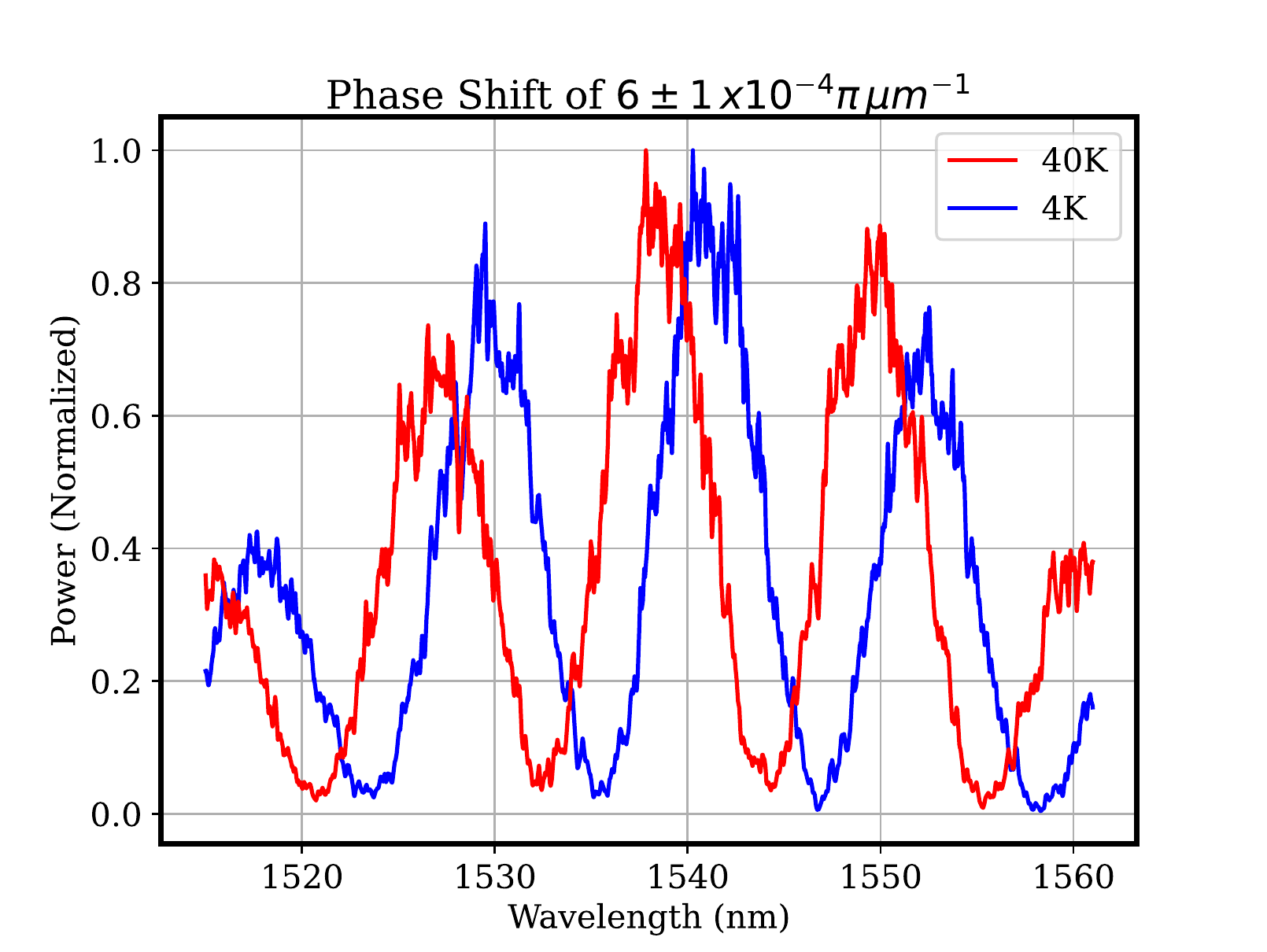}
    \caption{Showing stability of a 70~$nm$ Xe film when dropping from  to 4$K$. Phase shift of 6$\pm$1$\times$10$^{-4}$ $\pi/\mu m$ is the measured phase shift per micron of uncovered waveguide.}
    \label{fig:40K_4K_Stability}
\end{figure}

We can isolate the reasoning behind the behavior in Fig. S2 using Fig. S3. As the temperature drops from 200~$K$ to 40~$K$, we observe the increasing stability of the MZI as temperature decreases, with no shift observed between 75~$K$ and 40~$K$. This is before O$_2$ and N$_2$ can deposit on the sample, and suggests that a higher quality vacuum would almost certainly reduce the phase shift observed in Fig. S2.
Vacuum quality deteriorates in the presence of water vapor due to the non-zero humidity in our local environment, and the ability of water to bond strongly to the surfaces of exposed equipment \cite{BERMAN1996327}. We observe the effect of water vapor at much higher temperatures (Fig.~1a).
Preventative measures include leaving the vacuum chamber evacuated to desorb the majority of the water, in addition to keeping the sample temperature high during cool-down.\\

\begin{figure}[h]
    \centering
    \includegraphics[scale = 0.7]{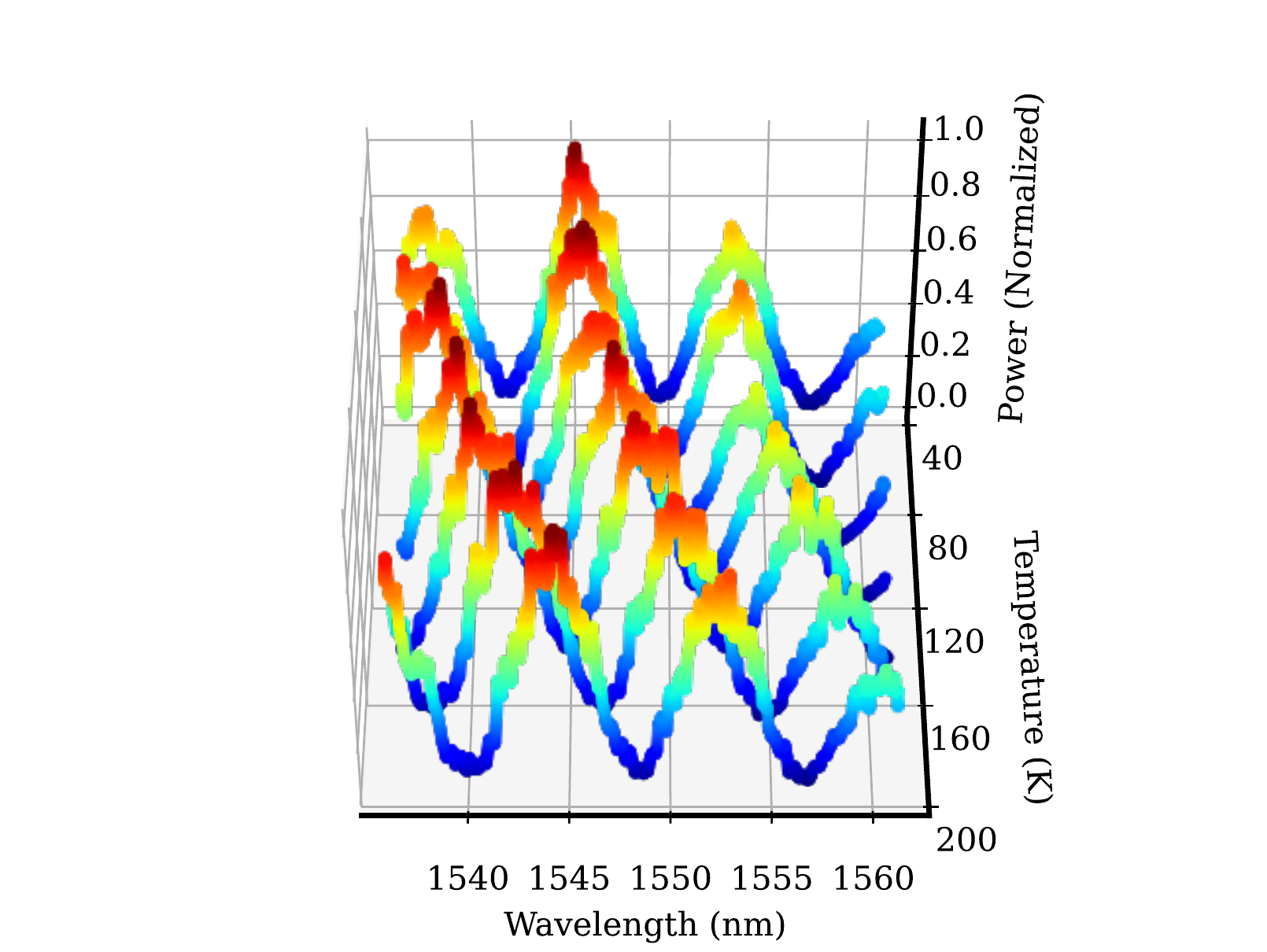}
    \caption{Showing stability of the uncovered MZI as temperature of the sample drops, demonstrating plateau of fringe shifting at 75$K$ - 40$K$. Here no Xe is present in the chamber, and all the radiation shields are at base temperature.}
    \label{fig:Temp_Drop_Stability}
\end{figure}

Another way to reduce the phase shift when dropping to 4~$K$ would be to use thicker films if a higher quality vacuum is not available. This is evidenced in Fig. S4 where we perform stability tests of over an hour in length. Figure S4a shows us the high stability expected from a thicker (110~$nm$) film, especially when compared to the original 70~$nm$ films (Fig. S4b) which were also used in our first point. Figure S4c shows the expected behavior of a vacuum system with a leak (data taken during late stage Xe depositions). This allows us to make the claim that even for non ultra-high vacuums, CLM can be used to create high-stability Xe films.\\

Finally, we can discuss some geometric considerations when it comes to a thermo-optic method of sublimating Xe films. Our heaters were located away from the unclad section of waveguide, to allow us to post-process the circuit. This resulted in some interesting behavior for higher heater powers, due to the lag time for heat to reach the majority of the Xe film. Figure S5(a-b) shows this as an addendum to Fig. 3(c-d), we can see that there is some asymmetry in the sublimation of the film, where the rate of sublimation rapidly increases after a sub-second period. This behavior can be further observed in Fig. 4c where some obvious asymmetry can be seen as the heaters are switched on. 

\begin{figure}[h]
    \centering
    \includegraphics[width=\linewidth]{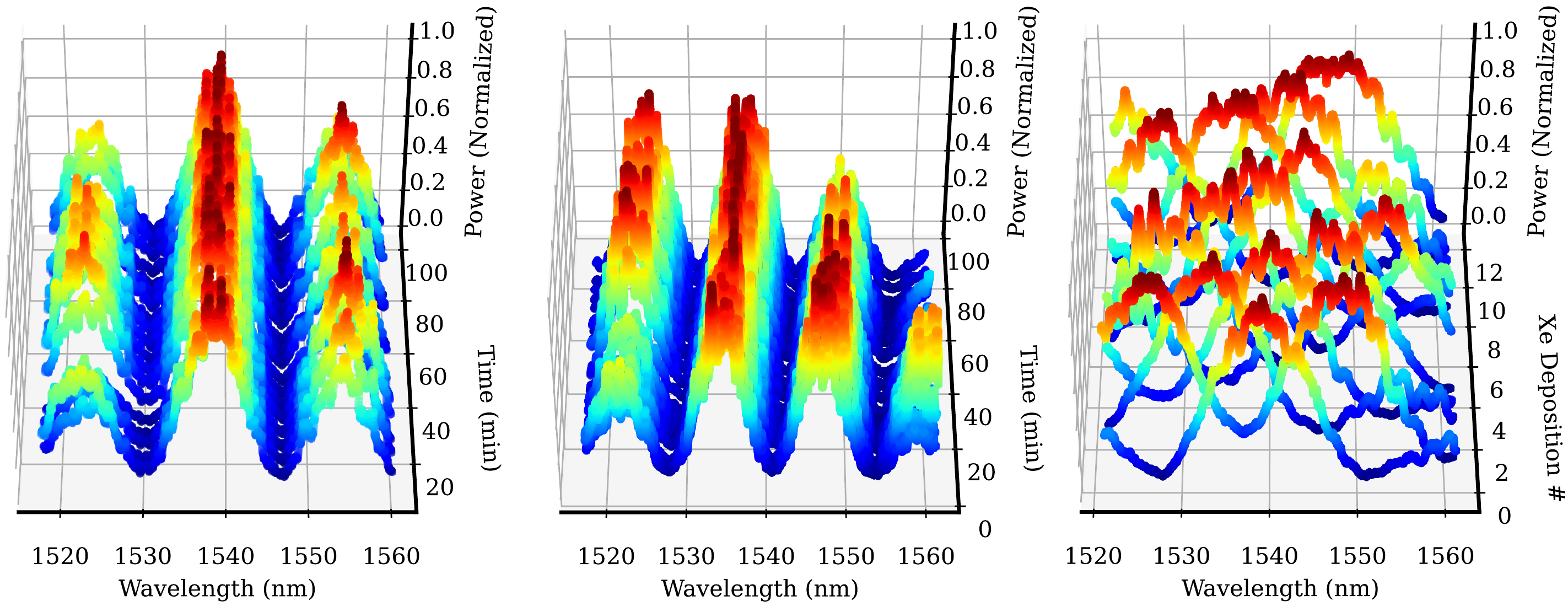}
    \caption{Plots showing stability of the Xe film in different scenarios. a) 110~$nm$ films, b) 70~$nm$ films, c) During Xe deposition (representative of a leaky system).}
    \label{fig:Thickness_Stability}
\end{figure}

Problems related to geometry can be alleviated in future implementations due to the small feature size required from a CLM device, minimizing lag times, and reducing the overall footprint. Alternatively, the heaters should be located in much closer proximity to the spiral, or another method of energy delivery is needed (such as LDW mentioned in the discussion). Existing devices (MZIs or otherwise) can also have their total device footprints reduced using Xe, by substituting or combining path mismatch with a change in $n_{g}$. Here power consumption is tied only to the changing of phase inside the device, steady-state configurations require no additional power to maintain.

\begin{figure}[h]
    \centering
    \includegraphics[scale = 1.1]{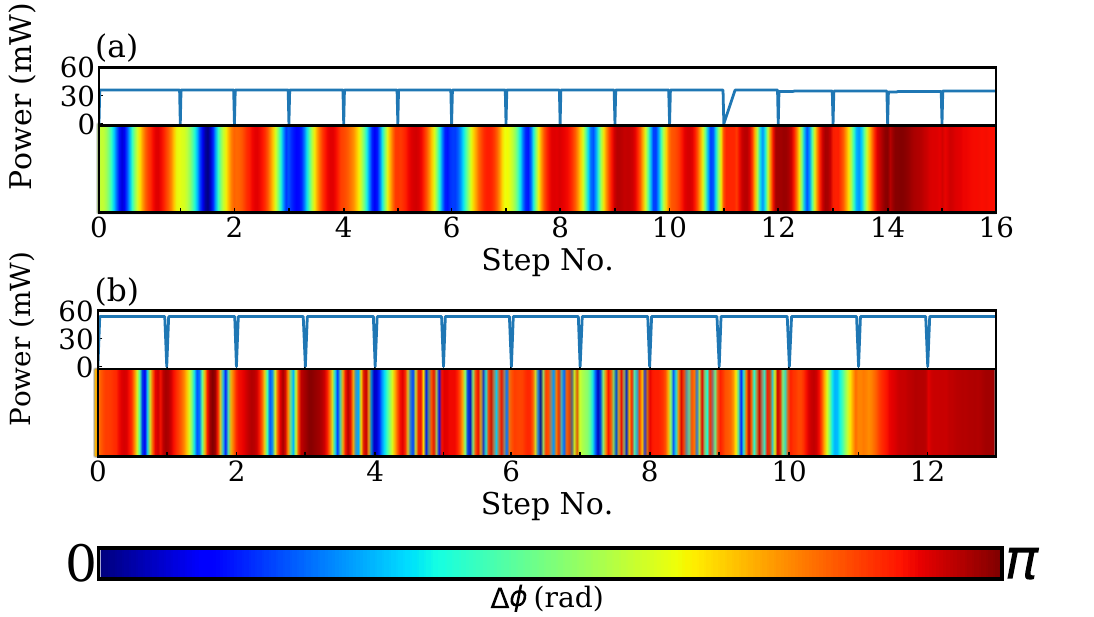}
    \caption{Plots (a, b) depict the electrical power delivered and the oscillation in relative phase ($\Delta\phi$) between the two arms of the MZI as the Xe sublimates with time as in Fig.~3.\\
    a) is from Fig.~3a. at 37$mW$ for 10$s$. b) is from Fig.~3b. at 55$mW$ for 10$s$. Here we notice some lag time in the phase shift of the interferometer which can be narrowed down to the geometry of the experiment.}
    \label{fig:Sublimation_Supp}
\end{figure}

\end{document}